\documentclass[aps,prl,superscriptaddress,twocolumn,amsmath,showkeys,showpacs]{revtex4-1}

\usepackage{units}
\usepackage{graphicx}
\usepackage{aas_macros}

\renewcommand{\vec}[1]{\mathbf{#1}}
\newcommand{\uvect}[1]{\mathbf{\hat{#1}}}

\begin{document}

\title{Probing Atmospheric Electric Fields in Thunderstorms through Radio Emission from Cosmic-Ray-Induced Air Showers}

\author{P.~Schellart}
\email[]{P.Schellart@astro.ru.nl}
\affiliation{Department of Astrophysics/IMAPP, Radboud University Nijmegen, PO Box 9010, 6500 GL Nijmegen, The Netherlands}
\author{T.~N.~G.~Trinh}
\affiliation{University of Groningen, KVI Center for Advanced Radiation Technology, 9700 AB Groningen, The Netherlands}
\author{S.~Buitink}
\affiliation{Astrophysical Institute, Vrije Universiteit Brussel, Pleinlaan 2, 1050 Brussels, Belgium}
\affiliation{Department of Astrophysics/IMAPP, Radboud University Nijmegen, PO Box 9010, 6500 GL Nijmegen, The Netherlands}
\author{A.~Corstanje}
\affiliation{Department of Astrophysics/IMAPP, Radboud University Nijmegen, PO Box 9010, 6500 GL Nijmegen, The Netherlands}
\author{J.~E.~Enriquez}
\affiliation{Department of Astrophysics/IMAPP, Radboud University Nijmegen, PO Box 9010, 6500 GL Nijmegen, The Netherlands}
\author{H.~Falcke}
\affiliation{Department of Astrophysics/IMAPP, Radboud University Nijmegen, PO Box 9010, 6500 GL Nijmegen, The Netherlands}
\affiliation{Nikhef, Science Park Amsterdam, 1098 XG Amsterdam, The Netherlands}
\affiliation{ASTRON, Netherlands Institute for Radio Astronomy, Postbus 2, 7990 AA Dwingeloo, The Netherlands}
\affiliation{Max-Planck-Institut f\"{u}r Radioastronomie, Auf dem H\"ugel 69, 53121 Bonn, Germany}
\author{J.~R.~H\"orandel}
\affiliation{Department of Astrophysics/IMAPP, Radboud University Nijmegen, PO Box 9010, 6500 GL Nijmegen, The Netherlands}
\affiliation{Nikhef, Science Park Amsterdam, 1098 XG Amsterdam, The Netherlands}
\author{A.~Nelles}
\affiliation{Department of Astrophysics/IMAPP, Radboud University Nijmegen, PO Box 9010, 6500 GL Nijmegen, The Netherlands}
\author{J.~P.~Rachen}
\affiliation{Department of Astrophysics/IMAPP, Radboud University Nijmegen, PO Box 9010, 6500 GL Nijmegen, The Netherlands}
\author{L.~Rossetto}
\affiliation{Department of Astrophysics/IMAPP, Radboud University Nijmegen, PO Box 9010, 6500 GL Nijmegen, The Netherlands}
\author{O.~Scholten}
\affiliation{University of Groningen, KVI Center for Advanced Radiation Technology, Groningen, The Netherlands}
\affiliation{Vrije Universiteit Brussel, Dienst ELEM, B-1050 Brussels, Belgium}
\author{S.~ter Veen}
\affiliation{Department of Astrophysics/IMAPP, Radboud University Nijmegen, PO Box 9010, 6500 GL Nijmegen, The Netherlands}
\affiliation{ASTRON, Netherlands Institute for Radio Astronomy, Postbus 2, 7990 AA Dwingeloo, The Netherlands}
\author{S.~Thoudam}
\affiliation{Department of Astrophysics/IMAPP, Radboud University Nijmegen, PO Box 9010, 6500 GL Nijmegen, The Netherlands}
\author{U.~Ebert}
\affiliation{Center for Mathematics and Computer Science (CWI), PO Box 94079, 1090 GB Amsterdam, The Netherlands}
\affiliation{Eindhoven University of Technology (TU/e), PO Box 513, 5600 MB Eindhoven, The Netherlands}
\author{C.~Koehn}
\affiliation{Center for Mathematics and Computer Science (CWI), PO Box 94079, 1090 GB Amsterdam, The Netherlands}
\author{C.~Rutjes}
\affiliation{Center for Mathematics and Computer Science (CWI), PO Box 94079, 1090 GB Amsterdam, The Netherlands}
\author{A.~Alexov}
\affiliation{Space Telescope Science Institute, 3700 San Martin Drive, Baltimore, Maryland 21218, USA}
\author{J.~M.~Anderson}
\affiliation{Helmholtz-Zentrum Potsdam, DeutschesGeoForschungsZentrum GFZ, Department 1: Geodesy and Remote Sensing, Telegrafenberg, A17, 14473 Potsdam, Germany}
\author{I.~M.~Avruch}
\affiliation{SRON Netherlands Institute for Space Research, PO Box 800, 9700 AV Groningen, The Netherlands}
\affiliation{Kapteyn Astronomical Institute, PO Box 800, 9700 AV Groningen, The Netherlands}
\author{M.~J.~Bentum}
\affiliation{ASTRON, Netherlands Institute for Radio Astronomy, Postbus 2, 7990 AA Dwingeloo, The Netherlands}
\affiliation{University of Twente, PO Box 217, 7500 AE Enschede, The Netherlands}
\author{G.~Bernardi}
\affiliation{Harvard-Smithsonian Center for Astrophysics, 60 Garden Street, Cambridge, Massachusetts 02138, USA}
\author{P.~Best}
\affiliation{Institute for Astronomy, University of Edinburgh, Royal Observatory of Edinburgh, Blackford Hill, Edinburgh EH9 3HJ, United Kingdom}
\author{A.~Bonafede}
\affiliation{University of Hamburg, Gojenbergsweg 112, 21029 Hamburg, Germany}
\author{F.~Breitling}
\affiliation{Leibniz-Institut f\"{u}r Astrophysik Potsdam (AIP), An der Sternwarte 16, 14482 Potsdam, Germany}
\author{J.~W.~Broderick}
\affiliation{Astrophysics, University of Oxford, Denys Wilkinson Building, Keble Road, Oxford OX1 3RH, United Kingdom}
\affiliation{School of Physics and Astronomy, University of Southampton, Southampton, SO17 1BJ, United Kingdom}
\author{M.~Br\"uggen}
\affiliation{University of Hamburg, Gojenbergsweg 112, 21029 Hamburg, Germany}
\author{H.~R.~Butcher}
\affiliation{Research School of Astronomy and Astrophysics, Australian National University, Mt. Stromlo Observatory, via Cotter Road, Weston, Australian Capital Territory 2611, Australia}
\author{B.~Ciardi}
\affiliation{Max Planck Institute for Astrophysics, Karl Schwarzschild Stra\ss e 1, 85741 Garching, Germany}
\author{E.~de Geus}
\affiliation{ASTRON, Netherlands Institute for Radio Astronomy, Postbus 2, 7990 AA Dwingeloo, The Netherlands}
\affiliation{SmarterVision BV, Oostersingel 5, 9401 JX Assen, The Netherlands}
\author{M.~de Vos}
\affiliation{ASTRON, Netherlands Institute for Radio Astronomy, Postbus 2, 7990 AA Dwingeloo, The Netherlands}
\author{S.~Duscha}
\affiliation{ASTRON, Netherlands Institute for Radio Astronomy, Postbus 2, 7990 AA Dwingeloo, The Netherlands}
\author{J.~Eisl\"offel}
\affiliation{Th\"{u}ringer Landessternwarte, Sternwarte 5, D-07778 Tautenburg, Germany}
\author{R.~A.~Fallows}
\affiliation{ASTRON, Netherlands Institute for Radio Astronomy, Postbus 2, 7990 AA Dwingeloo, The Netherlands}
\author{W.~Frieswijk}
\affiliation{ASTRON, Netherlands Institute for Radio Astronomy, Postbus 2, 7990 AA Dwingeloo, The Netherlands}
\author{M.~A.~Garrett}
\affiliation{ASTRON, Netherlands Institute for Radio Astronomy, Postbus 2, 7990 AA Dwingeloo, The Netherlands}
\affiliation{Leiden Observatory, Leiden University, PO Box 9513, 2300 RA Leiden, The Netherlands}
\author{J.~Grie\ss{}meier}
\affiliation{LPC2E - Univers\'ite d'Orleans/CNRS, 45071 Orleans Cedex 2, France}
\affiliation{Station de Radioastronomie de Nancay, Observatoire de Paris - CNRS/INSU, USR 704 - Univers\'ite Orleans, OSUC , Route de Souesmes, 18330 Nancay, France}
\author{A.~W.~Gunst}
\affiliation{ASTRON, Netherlands Institute for Radio Astronomy, Postbus 2, 7990 AA Dwingeloo, The Netherlands}
\author{G.~Heald}
\affiliation{ASTRON, Netherlands Institute for Radio Astronomy, Postbus 2, 7990 AA Dwingeloo, The Netherlands}
\affiliation{Kapteyn Astronomical Institute, PO Box 800, 9700 AV Groningen, The Netherlands}
\author{J.~W.~T.~Hessels}
\affiliation{ASTRON, Netherlands Institute for Radio Astronomy, Postbus 2, 7990 AA Dwingeloo, The Netherlands}
\affiliation{Anton Pannekoek Institute, University of Amsterdam, Postbus 94249, 1090 GE Amsterdam, The Netherlands}
\author{M.~Hoeft}
\affiliation{Th\"{u}ringer Landessternwarte, Sternwarte 5, D-07778 Tautenburg, Germany}
\author{H.~A.~Holties}
\affiliation{ASTRON, Netherlands Institute for Radio Astronomy, Postbus 2, 7990 AA Dwingeloo, The Netherlands}
\author{E.~Juette}
\affiliation{Astronomisches Institut der Ruhr-Universit\"{a}t Bochum, Universitaetsstrasse 150, 44780 Bochum, Germany}
\author{V.~I.~Kondratiev}
\affiliation{ASTRON, Netherlands Institute for Radio Astronomy, Postbus 2, 7990 AA Dwingeloo, The Netherlands}
\affiliation{Astro Space Center of the Lebedev Physical Institute, Profsoyuznaya Street 84/32, Moscow 117997, Russia}
\author{M.~Kuniyoshi}
\affiliation{National Astronomical Observatory of Japan, Tokyo 181-8588, Japan}
\author{G.~Kuper}
\affiliation{ASTRON, Netherlands Institute for Radio Astronomy, Postbus 2, 7990 AA Dwingeloo, The Netherlands}
\author{G.~Mann}
\affiliation{Leibniz-Institut f\"{u}r Astrophysik Potsdam (AIP), An der Sternwarte 16, 14482 Potsdam, Germany}
\author{R.~McFadden}
\affiliation{ASTRON, Netherlands Institute for Radio Astronomy, Postbus 2, 7990 AA Dwingeloo, The Netherlands}
\author{D.~McKay-Bukowski}
\affiliation{Sodankyl\"{a} Geophysical Observatory, University of Oulu, T\"{a}htel\"{a}ntie 62, 99600 Sodankyl\"{a}, Finland}
\affiliation{STFC Rutherford Appleton Laboratory, Harwell Science and Innovation Campus, Didcot OX11 0QX, United Kingdom}
\author{J.~P.~McKean}
\affiliation{ASTRON, Netherlands Institute for Radio Astronomy, Postbus 2, 7990 AA Dwingeloo, The Netherlands}
\affiliation{Kapteyn Astronomical Institute, PO Box 800, 9700 AV Groningen, The Netherlands}
\author{M.~Mevius}
\affiliation{ASTRON, Netherlands Institute for Radio Astronomy, Postbus 2, 7990 AA Dwingeloo, The Netherlands}
\affiliation{Kapteyn Astronomical Institute, PO Box 800, 9700 AV Groningen, The Netherlands}
\author{J.~Moldon}
\affiliation{ASTRON, Netherlands Institute for Radio Astronomy, Postbus 2, 7990 AA Dwingeloo, The Netherlands}
\author{M.~J.~Norden}
\affiliation{ASTRON, Netherlands Institute for Radio Astronomy, Postbus 2, 7990 AA Dwingeloo, The Netherlands}
\author{E.~Orru}
\affiliation{ASTRON, Netherlands Institute for Radio Astronomy, Postbus 2, 7990 AA Dwingeloo, The Netherlands}
\author{H.~Paas}
\affiliation{Center for Information Technology (CIT), University of Groningen, PO Box 72, 9700 AB Groningen, The Netherlands}
\author{M.~Pandey-Pommier}
\affiliation{Centre de Recherche Astrophysique de Lyon, Observatoire de Lyon, 9 Avenue Charles Andr\'{e}, 69561 Saint Genis Laval Cedex, France}
\author{R.~Pizzo}
\affiliation{ASTRON, Netherlands Institute for Radio Astronomy, Postbus 2, 7990 AA Dwingeloo, The Netherlands}
\author{A.~G.~Polatidis}
\affiliation{ASTRON, Netherlands Institute for Radio Astronomy, Postbus 2, 7990 AA Dwingeloo, The Netherlands}
\author{W.~Reich}
\affiliation{Max-Planck-Institut f\"{u}r Radioastronomie, Auf dem H\"ugel 69, 53121 Bonn, Germany}
\author{H.~R\"ottgering}
\affiliation{Leiden Observatory, Leiden University, PO Box 9513, 2300 RA Leiden, The Netherlands}
\author{A.~M.~M.~Scaife}
\affiliation{School of Physics and Astronomy, University of Southampton, Southampton, SO17 1BJ, UK}
\author{D.~J.~Schwarz}
\affiliation{Fakult\"{a}t f\"{u}r Physik, Universit\"{a}t Bielefeld, Postfach 100131, D-33501, Bielefeld, Germany}
\author{M.~Serylak}
\affiliation{Astrophysics, University of Oxford, Denys Wilkinson Building, Keble Road, Oxford OX1 3RH}
\author{O.~Smirnov}
\affiliation{Department of Physics and Electronics, Rhodes University, PO Box 94, Grahamstown 6140, South Africa}
\affiliation{SKA South Africa, 3rd Floor, The Park, Park Road, Pinelands, 7405, South Africa}
\author{M.~Steinmetz}
\affiliation{Leibniz-Institut f\"{u}r Astrophysik Potsdam (AIP), An der Sternwarte 16, 14482 Potsdam, Germany}
\author{J.~Swinbank}
\affiliation{Anton Pannekoek Institute, University of Amsterdam, Postbus 94249, 1090 GE Amsterdam, The Netherlands}
\author{M.~Tagger}
\affiliation{LPC2E - Universite d'Orleans/CNRS, 45071 Orleans cedex 2, France}
\author{C.~Tasse}
\affiliation{LESIA, UMR CNRS 8109, Observatoire de Paris, 92195 Meudon, France}
\author{M.~C.~Toribio}
\affiliation{ASTRON, Netherlands Institute for Radio Astronomy, Postbus 2, 7990 AA Dwingeloo, The Netherlands}
\author{R.~J.~van Weeren}
\affiliation{Harvard-Smithsonian Center for Astrophysics, 60 Garden Street, Cambridge, MA 02138, USA}
\author{R.~Vermeulen}
\affiliation{ASTRON, Netherlands Institute for Radio Astronomy, Postbus 2, 7990 AA Dwingeloo, The Netherlands}
\author{C.~Vocks}
\affiliation{Leibniz-Institut f\"{u}r Astrophysik Potsdam (AIP), An der Sternwarte 16, 14482 Potsdam, Germany}
\author{M.~W.~Wise}
\affiliation{ASTRON, Netherlands Institute for Radio Astronomy, Postbus 2, 7990 AA Dwingeloo, The Netherlands}
\affiliation{Anton Pannekoek Institute, University of Amsterdam, Postbus 94249, 1090 GE Amsterdam, The Netherlands}
\author{O.~Wucknitz}
\affiliation{Max-Planck-Institut f\"{u}r Radioastronomie, Auf dem H\"ugel 69, 53121 Bonn, Germany}
\author{P.~Zarka}
\affiliation{LESIA, UMR CNRS 8109, Observatoire de Paris, 92195   Meudon, France}

\date{\today}

\begin{abstract}
We present measurements of radio emission from cosmic ray air showers that took place during thunderstorms. The intensity and polarization patterns of these air showers are radically different from those measured during fair-weather conditions. With the use of a simple two-layer model for the atmospheric electric field, these patterns can be well reproduced by state-of-the-art simulation codes. This in turn provides a novel way to study atmospheric electric fields.
\end{abstract}

\pacs{92.60.Pw, 95.85.Ry, 96.50.sd}
\keywords{cosmic rays; thunderstorms}

\maketitle

One of the important open questions in atmospheric physics concerns the physical mechanism that initiates lightning in thunderclouds \citep{Dwyer:2014}. Crucial to the answer is knowledge of atmospheric electric fields. Existing \emph{in situ} measurements, from balloons or airplanes, are limited due to the violent nature of thunderstorms. Furthermore, they are limited to balloon trajectories or perturbed by the presence of the aircraft. Here we present a new method to probe atmospheric electric fields through their influence on the pattern of polarized radio emission emitted by cosmic-ray-induced extensive air showers.

The main mechanism for driving radio-wave emission from air showers is that the relativistic electrons and positrons in the electromagnetic part of the shower are accelerated in opposite directions by the Lorentz force exerted by Earth's magnetic field. This produces a short, nanosecond time scale, coherent pulse of radio emission mostly at megahertz frequencies. The emission generated by this geomagnetic mechanism is unidirectionally polarized in the $\uvect{e}_{\vec{v}\times\vec{B}}$ direction. Here, $\vec{v}$ is the propagation velocity vector of the shower and $\vec{B}$ represents Earth's magnetic field \citep{Kahn:1966,Falcke:2005,Codalema2009}.

A secondary emission mechanism, contributing between $\sim3{-}20\%$ to the signal amplitude depending on distance to the shower axis and the arrival direction of the shower \citep{Aab:2014,Schellart:2014}, results from a negative charge excess in the shower front. This consists of electrons knocked out of air molecules by the air shower. This also produces a short radio pulse but now polarized radially with respect to the shower symmetry axis.

The emission from both processes is strongly beamed in the forward direction, due to the relativistic velocities of the particles. Additionally, the nonunity refractive index of the air causes relativistic time-compression effects leading to enhanced emission from parts of the shower seen at the Cherenkov angle \citep{Vries:2011,Nelles:2014}. Interference between the differently polarized emission from both components leads to a complex and highly asymmetric \emph{intensity pattern} \citep{de-Vries:2010}. In contrast, time-compression effects do not alter the direction of the polarization vector of the emission. The \emph{polarization pattern} of the radio emission thus points predominantly in the $\uvect{e}_{\vec{v}\times\vec{B}}$ direction with a minor radial deviation. Strong atmospheric electric fields will influence the motions of the electrons and positrons in air showers. This is expected to be visible in the polarization patterns of the recorded emission \citep{Buitink:2010}. Therefore we analyze air showers recorded during thunderstorms.

Data for this analysis were recorded with the low-band, $\unit[10{-}90]{MHz}$, dual-polarized crossed dipole antennas located in the inner, $\sim\unit[2]{km}$ radius, core of the LOw-Frequency ARray (LOFAR) radio telescope \citep{van-Haarlem:2013}. These antennas are grouped into circular stations that act as dishes for standard interferometric astronomical observations. For the purpose of air shower measurements, all antennas are equipped with ring buffers that can store up to $\unit[5]{s}$ of raw voltage data sampled every $\unit[5]{ns}$. A dedicated scintillator array, LOfar Radboud air shower Array (LORA), is located at the center of LOFAR to provide an independent trigger whenever an air shower with an estimated primary energy of $\ge\unit[2\times 10^{16}]{eV}$ is detected \citep{Thoudam:2014}. When a trigger is received, $\unit[2]{ms}$ of raw voltage data around the trigger time are stored for every active antenna.

These data are processed in an off-line analysis \citep{Schellart:2013} from which a number of physical parameters are extracted and stored. These include the estimated energy of the air shower (as reconstructed from the particle detector data), the arrival direction of the air shower (as reconstructed from the arrival times of the radio pulses in all antennas), and for each antenna polarization information in the form of the Stokes parameters: I (intensity), Q, U and V. The orientation of the polarization vector is reconstructed from Stokes Q and U \citep{Schellart:2014}.

Over the period between June 2011 and September 2014, LOFAR recorded a total of $762$ air showers. The complex radio intensity pattern on the ground of almost all measured showers can be well reproduced by state-of-the-art air shower simulation codes \citep{Buitink:2014a}. These codes augment well-tested Monte Carlo air shower simulations with radio emission calculated from first principles at the microscopic level \citep{Huege:2013,Alvarez-Muniz:2012}. In this analysis, we use the CoREAS plugin of CORSIKA \citep{Heck:1998} with QGSJETII \citep{Ostapchenko:2006} and FLUKA \citep{Battistoni:2007} as the hadronic interaction models. It was previously found that the exact shape of the intensity pattern depends on the atmospheric depth where the number of shower particles is largest, $X_\mathrm{max}$, and that the absolute field strength of the radio emission scales with the energy of the primary particle.

\begin{figure}
\centering
\includegraphics[width=0.44\textwidth]{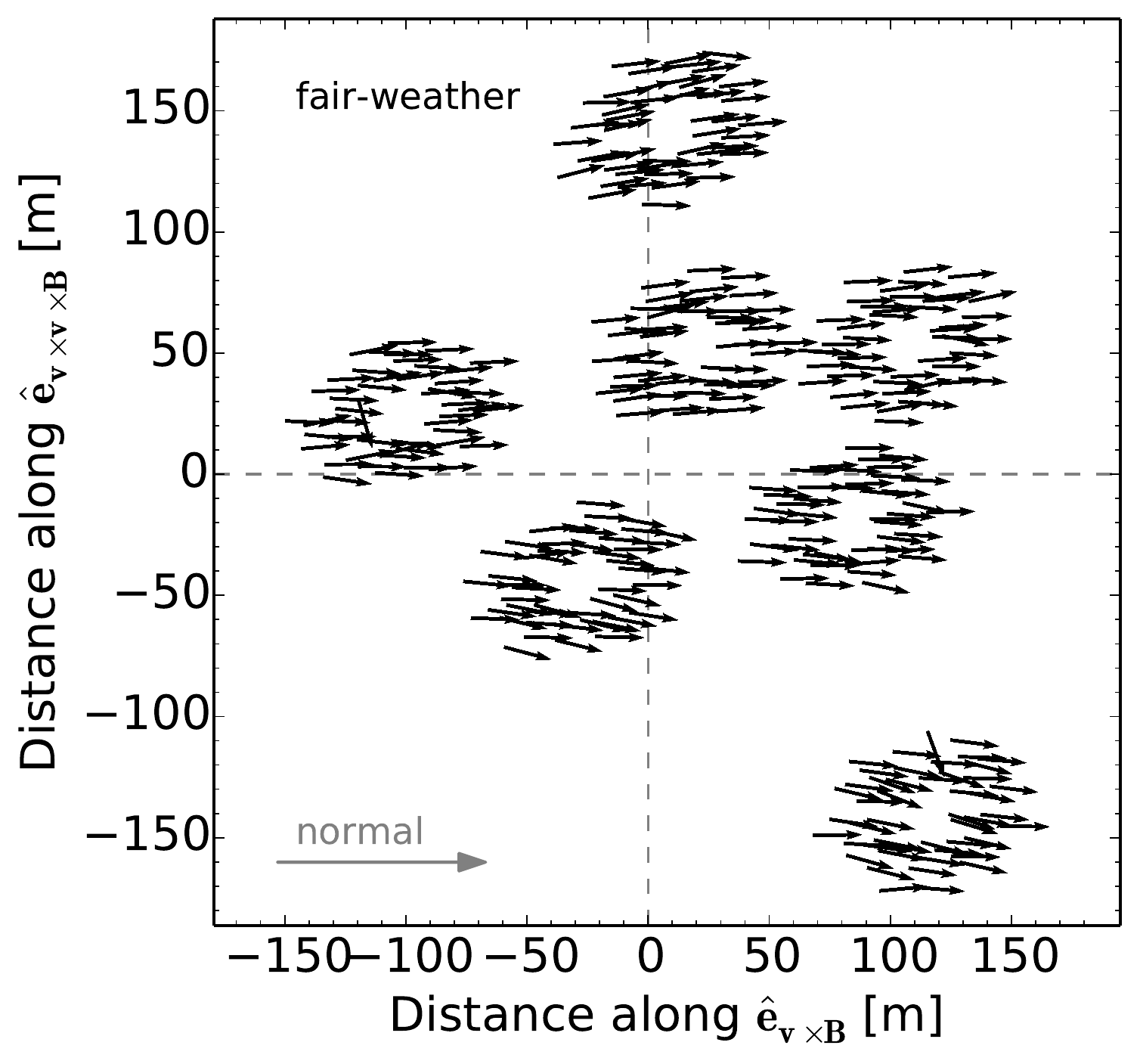}\\
\includegraphics[width=0.44\textwidth]{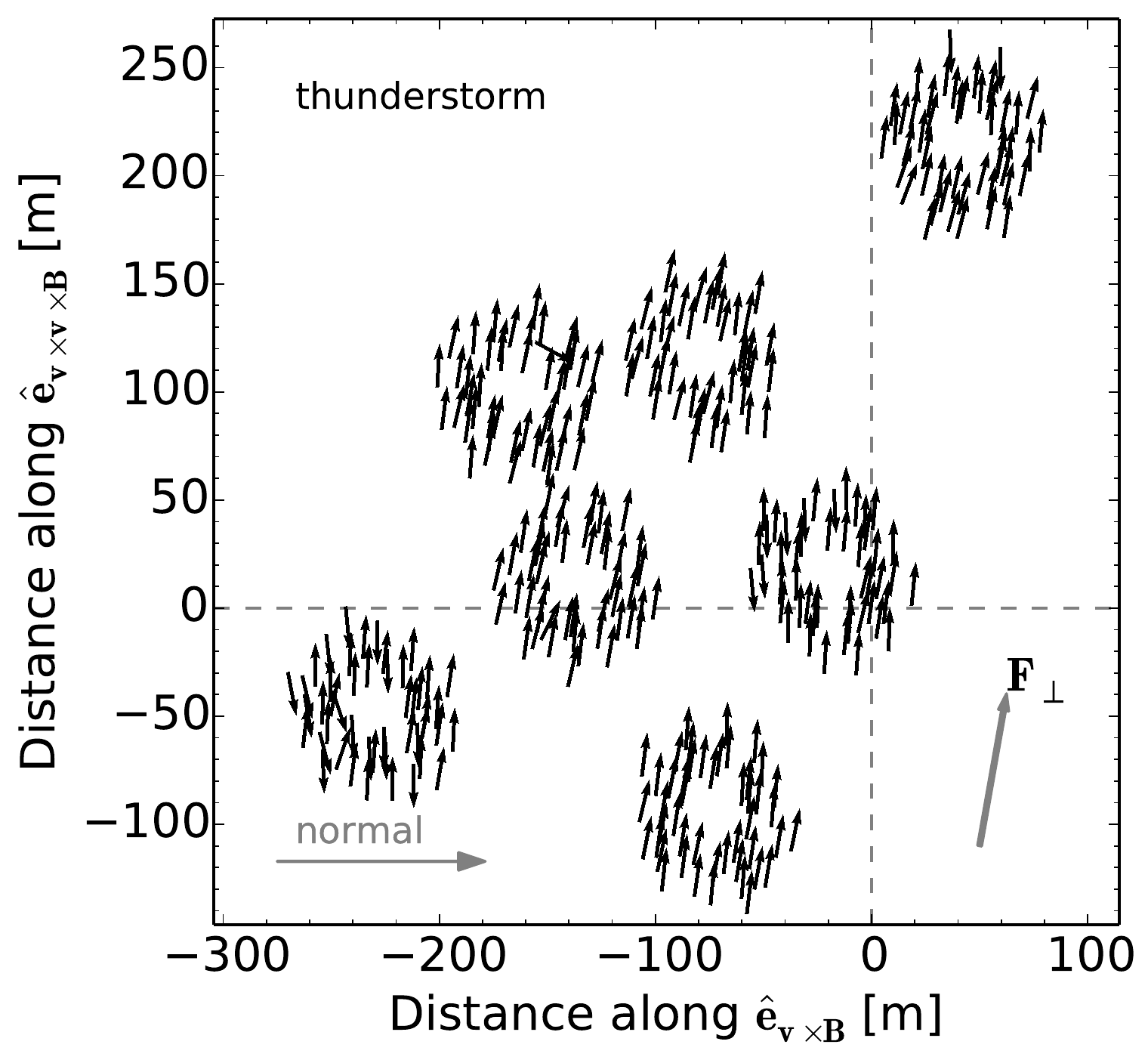}\\
\includegraphics[width=0.44\textwidth]{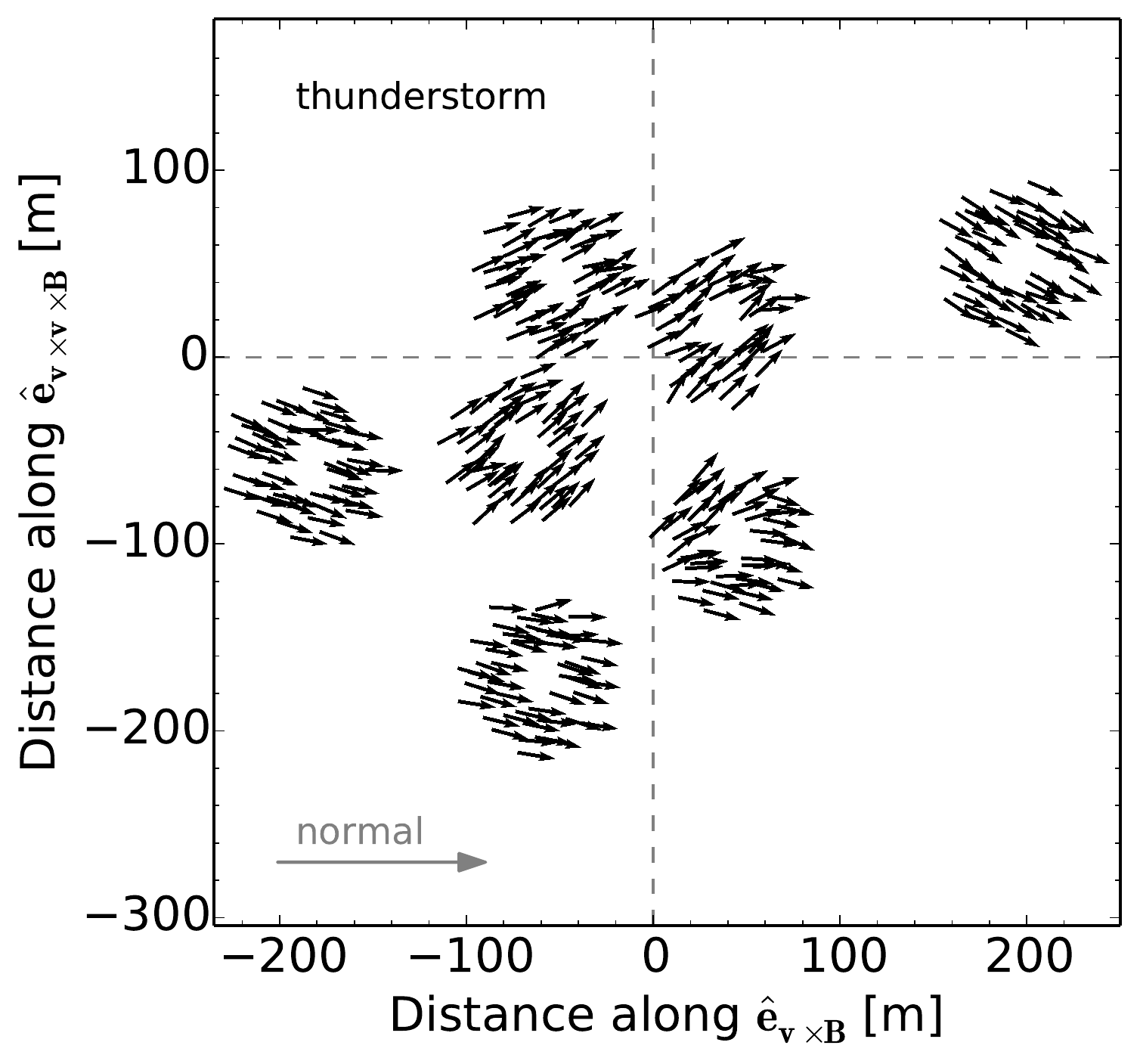}
\caption{Polarization as measured with individual LOFAR antennas (arrows) in the shower plane for three measured air showers. LOFAR antennas are grouped into circular stations, of which seven are depicted. The expected polarization direction for fair-weather air showers is indicated with ``normal''. The position of the shower axis, orthogonal to the shower plane, is indicated by the intersection of the dashed lines.}
\label{fig:polarization_footprints}
\end{figure}

\begin{figure}
\centering
\includegraphics[width=0.45\textwidth]{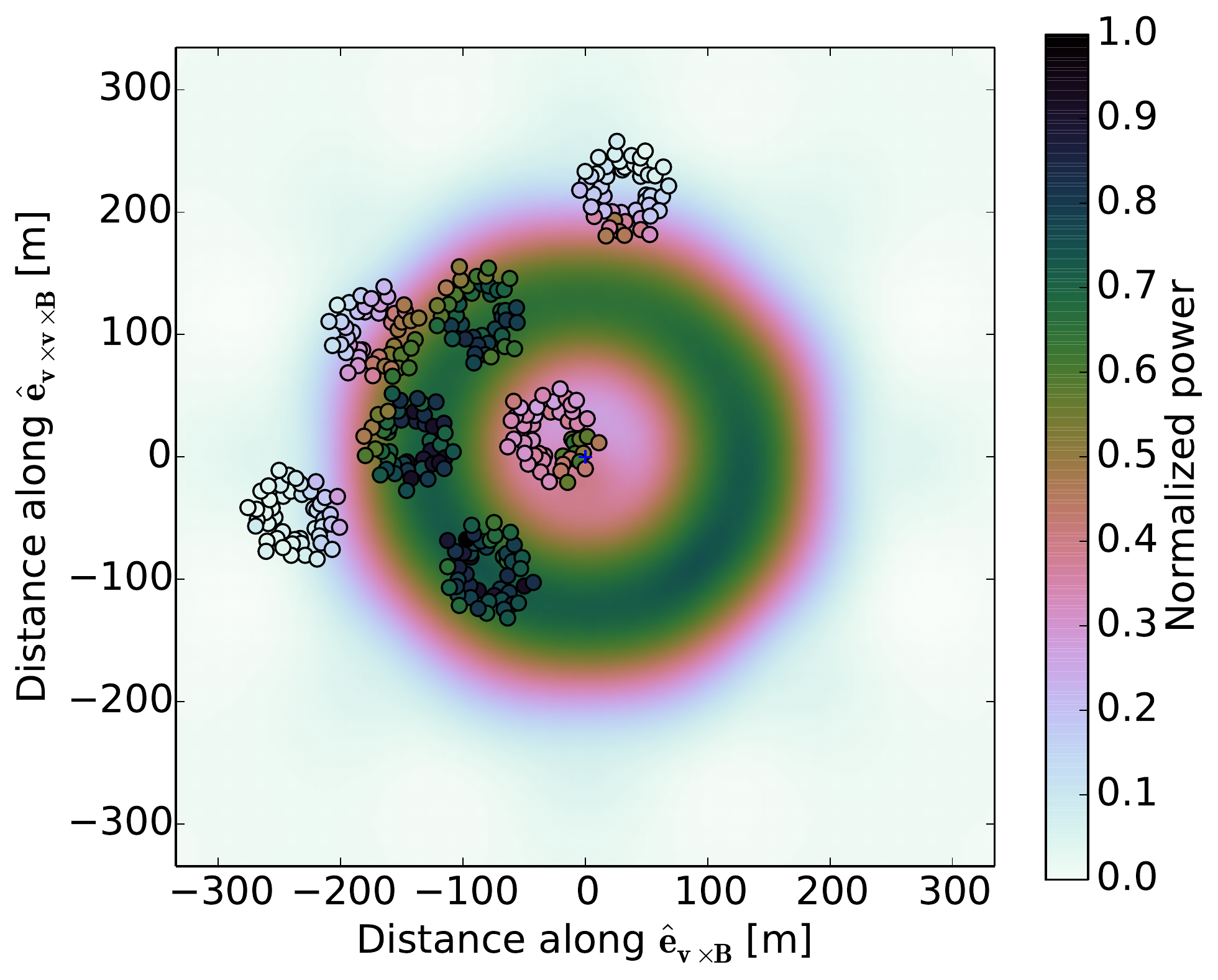}\\
\includegraphics[width=0.45\textwidth]{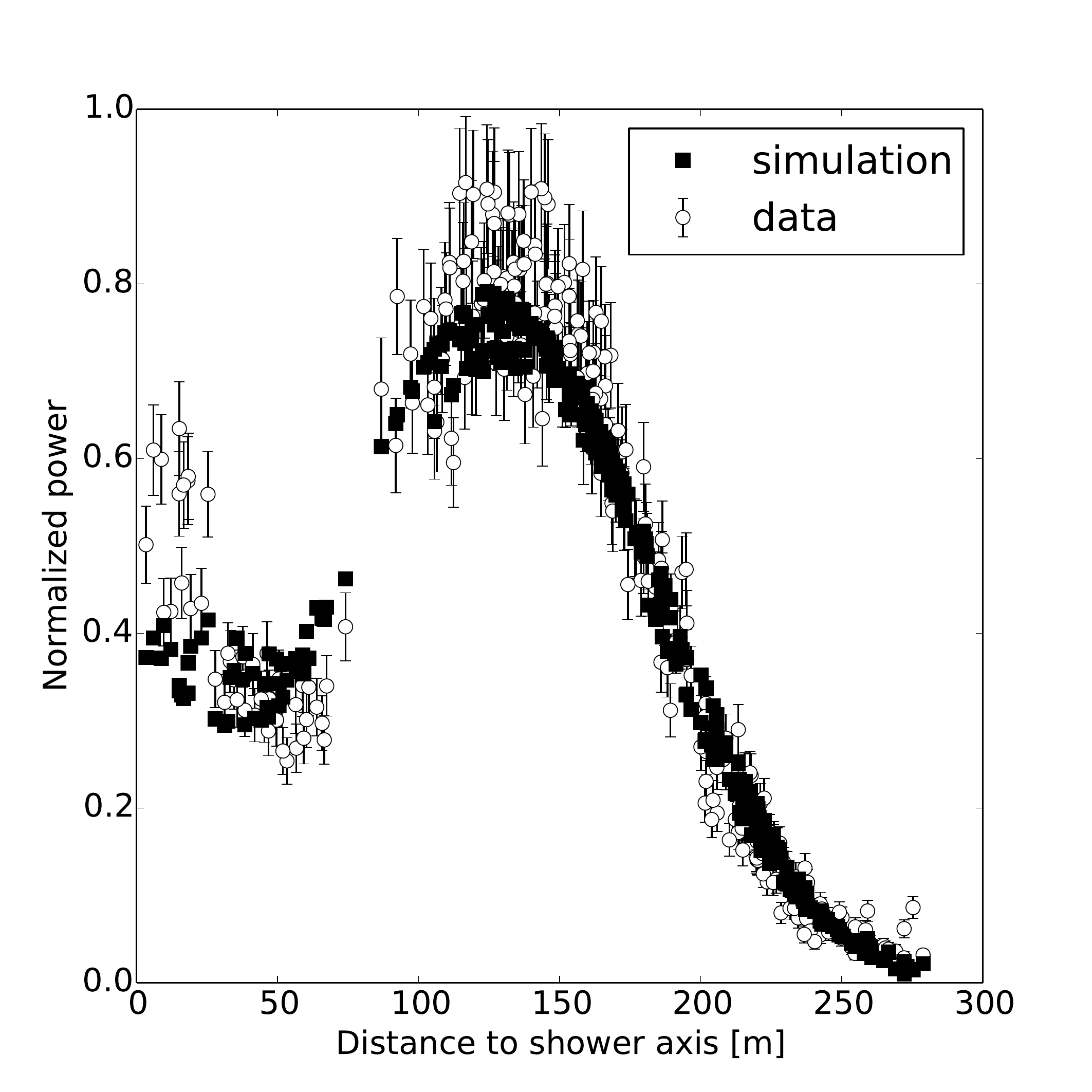}
\caption{Radio intensity pattern during a thunderstorm. Top: the circles represent antenna positions. Their color reflects measured pulse power. The best-fitting CoREAS simulation is shown in color scale in the background. Where the colors of the circles match the background a good fit is achieved. Bottom: measured (circles) and simulated pulse power (squares) as a function of distance to the shower axis.}
\label{fig:example_thunderstorm_event}
\end{figure}

The radio footprints of $58$ of the $762$ air showers are very different from those predicted by simulations. Of these, $27$ air showers have a measured signal-to-noise ratio below $10$ in amplitude --- too low to get a reliable reconstruction. The polarization patterns of the other $31$ showers differ significantly from those of ``normal'' fair-weather air showers. This can be seen in the middle and bottom panels of Fig.~\ref{fig:polarization_footprints} where the polarization direction is clearly coherent (i.e., nonrandom) over all antennas but no longer in the expected $\uvect{e}_{\vec{v}\times\vec{B}}$ direction.
In addition, for some of these showers the intensity of the radio signal at low $\unit[10{-}90]{MHz}$ frequencies is strongest on a ring around the shower axis with a radius of approximately $\unit[100]{m}$ (see also Fig.~\ref{fig:example_thunderstorm_event}). This ``ring structure'' in the intensity pattern is not present in normal fair-weather air showers that all lack rotational symmetry in the intensity pattern and instead show a single maximum that is displaced in the $\uvect{e}_{\vec{v}\times\vec{B}}$ direction from the shower axis \citep{Buitink:2014a,Nelles:2015}.
Twenty of these $31$ showers occur within $\unit[2]{h}$ of lightning strikes recorded within $\sim\unit[150]{km}$ distance from LOFAR by the Royal Dutch Meteorological Institute. Given the similarity of the polarization patterns of the remaining showers where no lightning strikes were measured, it is plausible that at these times the atmospheric electric field was also strong albeit not strong enough to initiate lightning. An electric field meter has since been installed at LOFAR that will provide independent verification for future measurements.

For the shower in the middle panel of Fig.~\ref{fig:polarization_footprints}, recorded during thunderstorm conditions, the pattern is unidirectional for the entire footprint. A second more complicated type is depicted in the bottom panel. Here, the pattern is more ``wavy''. The analysis presented here focuses on an air shower of the first type where also a ring structure is visible and a strong signal is measured by the LORA particle detectors. All air showers of this type can be reconstructed with similar accuracy. For showers of the wavy type a more complex analysis is currently being developed.

We propose that the influence of atmospheric electric fields on air shower radio emission can be understood in the following way.

The electric field, in the region of the cloud traversed by the air shower, can be decomposed into components perpendicular $\vec{E}_{\perp}$ and parallel $\vec{E}_{\parallel}$ to the shower symmetry axis. The perpendicular component of the field changes the net transverse force acting on the particles
\begin{equation}
\vec{F} = q(\vec{E}_{\perp} + \vec{v}\times\vec{B}).
\end{equation}
This changes both the magnitude and the polarization of the radiation that follow $\vec{F}$.

During shower development the air shower particles lose energy. The parallel component of the atmospheric electric field partially compensates this energy loss. Therefore, the total number of particles within a given energy range in the shower increases. Because the fractional gain of energy is greatest for lower energy particles, these are the most affected. However, low-energy particles do not contribute much to the total radio emission because they lag behind the shower front and their emission is not coherent for frequencies above $\unit[10]{MHz}$. Thus, it is the perpendicular component of the electric field that determines the measured intensity and polarization direction.

In order to test these hypotheses, atmospheric electric fields were inserted into CoREAS air shower simulations. By the comparison of fields acting purely parallel and purely perpendicular to the shower axis it was found that the effect of $\vec{E}_{\perp}$ on the radio emission is indeed much stronger and will dominate in most shower configurations where both components are present. This will be discussed in greater detail in a forthcoming publication.

Having understood the basic effects of atmospheric electric fields on air shower radio emission we proceed with a full reconstruction of LOFAR measurements. We follow the method developed by \citet{Buitink:2014a} to fit CoREAS simulations to LOFAR measurements. An atmospheric electric field is inserted into the simulations with the perpendicular component chosen such that the net force is in the measured average polarization direction (as indicated in the middle panel of Fig.~\ref{fig:polarization_footprints}). The parallel component is set to zero since its influence on the received radiation intensity and polarization pattern is negligible.

The simplest electric field configuration that can reproduce the main features in both the measured intensity and polarization patterns is composed of two electric field layers. The upper layer, with strength $|\vec{E}_{U}|$, starts at a height $h_U$ above the ground and extends down to a height $h_L$ at which the lower layer starts, the direction of the net force changes by $180^\circ$ and the field strength decreases to $|\vec{E}_{L}|$. Two layers are needed because with one layer the ring structure seen in the measurements is not reproducible.

In Fig.~\ref{fig:example_thunderstorm_event} the reconstruction is shown for the air shower for which the polarization pattern is depicted in the middle panel of Fig.~\ref{fig:polarization_footprints}. The reconstruction is optimal for $h_U=\unit[8]{km}$, $h_L=\unit[2.9]{km}$, $|\vec{E}_U|=\unit[50]{kV\, m^{-1}}$ and $|\vec{E}_L|/|\vec{E}_U|=0.53$. For these values $\chi^2/\mathrm{ndf}=3.2$ as obtained for a joined fit to both the radio and particle data. A perfect fit of $\chi^2/\mathrm{ndf}\approx 1$, as is often found for fair-weather showers, is likely not attainable with a simplified electric field model. However, all the main features of the intensity and polarization pattern (namely the overall polarization direction and ring structure) are already correctly reproduced.

\begin{figure*}
\centering
\includegraphics[width=1.0\textwidth]{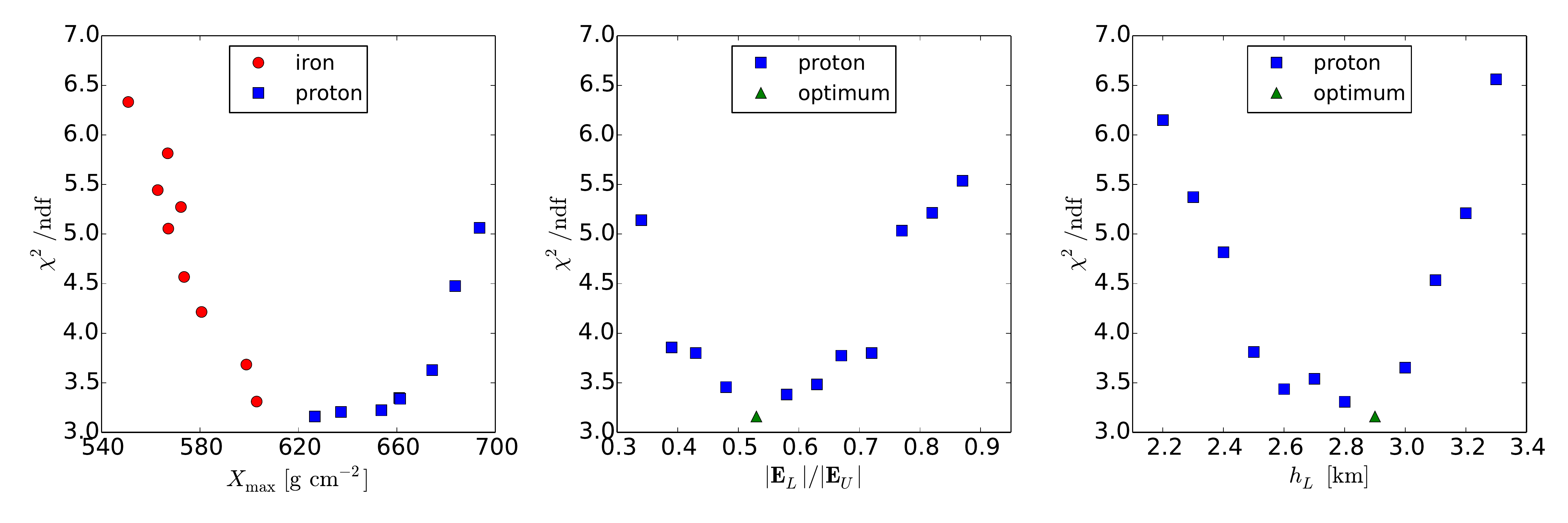}
\caption{Sensitivity of the fit quality to variations in the atmospheric depth of shower maximum $X_\mathrm{max}$ (left panel), the relative field strength (middle panel) and the field reversal altitude $h_2$ (right panel). The optimal proton simulation is the same for all plots. The electric field strength, in the upper layer, for all simulations is $|\vec{E}_{U}|=\unit[50]{kV\, m^{-1}}$.}
\label{fig:chi2_mapping}
\end{figure*}

The fit quality is sensitive to changes in the relative field strength and $h_L$ as well as $X_\mathrm{max}$. This can be seen in Fig.~\ref{fig:chi2_mapping}, where each parameter is varied while keeping the others fixed at their optimum values. This fixing is not possible for $X_\mathrm{max}$ in the CORSIKA software, because it is a an outcome of the simulation rather than an input parameter. Therefore, simulations were selected where $X_\mathrm{max}$ differs by no more than $\unit[20]{g\, cm^{-2}}$. The fit quality reaches its optimum value for $h_U=\unit[8]{km}$ and is not sensitive to a further increase. This is expected because above this altitude the air shower is not yet fully developed and there are relatively few particles contributing to the emission.

For fair-weather air showers the measured radio intensity is related to the simulated values through a constant scaling factor \citep{Buitink:2014a} given the energy of the primary particle. This energy is derived from the particle density on the ground, as measured with LORA, combined with the information on $X_\mathrm{max}$, as determined from the radio fit. For the air shower measured during thunderstorm conditions the measured intensity is higher than the normally expected value, as the absolute electric field strength influences the radio intensity. However, the simulated intensity increases only until the atmospheric electric field strength reaches $|\vec{E}_U|\geq\unit[50]{kV\, m^{-1}}$. When the field strength is increased further the radio intensity stays constant. This saturation of the radio intensity appears to be related to the coherent nature of the emission but is still under investigation.

Measuring radio emission from extensive air showers during thunderstorm conditions thus provides a unique new tool to probe the atmospheric electric fields present in thunderclouds. Unlimited by violent wind conditions and sensitive to a large fraction of the cloud this technique may help answer the long-standing question ``how is lighting initiated in thunderclouds?'' It has been suggested by Gurevich \textit{et al.}~\cite{Gurevich:1999,Gurevich:2013} that cosmic-ray-induced air showers in combination with runaway breakdown may initiate lightning. If this is indeed true then LOFAR with its combination of particle detectors and radio antennas is well positioned to measure it.

\begin{acknowledgments}
The LOFAR key science project cosmic rays acknowledges financial support from NOVA, SNN and FOM as well as from NWO, VENI Grant No. 639-041-130. We acknowledge funding from an Advanced Grant of the European Research Council (FP/2007-2013) / ERC Grant Agreement No. 227610.
We furthermore acknowledge funding from the FOM-project 12PR3041 ``Cosmic Lightning''.
LOFAR, the Low Frequency Array designed and constructed by ASTRON, has facilities in several countries that are owned by various parties (each with their own funding sources), and that are collectively operated by the International LOFAR Telescope foundation under a joint scientific policy.
\end{acknowledgments}

\bibliography{thunderstorms}

\end{document}